\documentclass[pre, aps,twocolumn]{revtex4}
\usepackage{amsfonts,amssymb,amsmath,bm}

\begin{document}

\title{Structure of coherent vortices generated by the inverse cascade of $2d$ turbulence in a finite box}

\author{I.V.Kolokolov and V.V.Lebedev}

\affiliation{Landau Institute for Theoretical Physics, RAS, \\
142432, Chernogolovka, Moscow region, Russia}

\begin{abstract}

We discuss structure and geometrical characteristics of coherent vortices appearing as a result of the inverse cascade in the two-dimensional ($2d$) turbulence in a finite box. We demonstrate that the universal velocity profile, established in \cite{14LBFKL}, corresponds to the passive regime of flow fluctuations. We find the vortex core radius and the vortex size and argue that an amount of the vortices generated in the box depends on the system parameters.

\end{abstract}


\maketitle

\section{Introduction}

A role of the counteraction of turbulence fluctuations with a mean (coherent) flow is one of the central problems of turbulence theory  \cite{townsend80}. Usually the fluid energy is transferred from the large-scale flow to turbulent pulsations \cite{Frisch}. However, in some cases the energy can go from small-scale fluctuations to the large-scale ones that can lead to formation of a mean flow \cite{BE12}. Even basic problems such as to determine at which mean velocity turbulent fluctuations are sustained is still object of intense investigations \cite{avila2011}. There is still no consistent theory for the mean (coherent) profile coexisting with turbulent fluctuations, so that even the celebrated logarithmic law for the turbulent boundary layer is a subject of controversy \cite{deb}. Here, we consider $2d$ turbulence in a restricted box where large-scale coherent structures are generated from small-scale fluctuations excited by pumping. This process occurs because in $2d$ the non-linear interaction favors the energy transfer to larger scales \cite{67Kra,68Lei,69Bat}.

Already, the first experiments on $2d$ turbulence \cite{Som} have shown that in a finite system with small bottom friction, the energy transfer to large scales leads to the formation of coherent vortices. Numerical simulations \cite{Smith1,Smith2,Borue} also show appearing coherent vortices in $2d$ turbulence. Subsequent numerical simulations \cite{Colm} and experiments \cite{Shats} demonstrated that these vortices have well-defined and reproducible mean velocity (vorticity) profiles. This profile is quite isotropic with a power-law radial decay of vorticity inside the vortex. The profile in that region depends neither on the boundary conditions (no-slip in experiments, periodic in numerics) nor on the type of forcing (random in numerics versus parametric excitation or electromagnetic force in experiments). The same profile is formed both in the statistically stationary case where the mean flow level is stabilized by the bottom friction and in the case where the average flow is still not stabilized and increases as time runs.

In the paper \cite{14LBFKL} results of intensive simulations of $2d$ turbulence were reported, they demonstrated that the vortex polar velocity profile is flat in some interval of distances from the vortex center. That means that the average vorticity is inversely proportional to the distance $r$ from the vortex center. In the same paper theoretical scheme based on conservation laws and on symmetry arguments was proposed that explains the flat velocity profile. The scheme predicts the value of the polar velocity $U=\sqrt{3\epsilon/\alpha}$ (where $\epsilon$ is the energy production rate and $\alpha$ is the bottom friction coefficient), that is in excellent agreement with the numerics \cite{14LBFKL}. However, in the numerics the region of existence of the flat profile is definitely restricted. In addition, in early simulations \cite{Smith1,Smith2,Borue} no flat velocity profile was observed. The facts needs an explanation.

That is why we performed a detailed analytical investigation of the problem of $2d$ turbulence in a finite box. As a result, we established that the flat velocity profile corresponds to the passive regime of  the flow fluctuations where their self-interaction can be neglected. The passive regime admits consistent analytical calculations that confirm validity of the value $U=\sqrt{3\epsilon/\alpha}$ for the polar velocity. Besides, we found expressions for the viscous core radius of the vortex and for the border of the region where the flat velocity profile is realized. The results explain why no flat velocity profile was observed in early simulations \cite{Smith1,Smith2,Borue} and imply that at some conditions a large number of coherent vortices could appear instead of a few vortices in numerics \cite{14LBFKL,Colm} and experiment \cite{Shats}.

\section{General Relations}
\label{sec:general}

We consider the case where $2d$ turbulence is excited in a finite box of size $L$ by external forcing. It is assumed to be random with homogeneous in time and space statistical properties. We assume also that the properties are isotropic. The main object of our investigation is the stationary (in the statistical sense) turbulent state that is caused by such forcing.

For exciting turbulence the forcing should be stronger than dissipation that is caused both by bottom friction and viscosity. That implies that the characteristic velocity gradient of the fluctuations produced by the forcing should be much larger than their damping at the pumping scale. The velocity gradient is estimated as $\epsilon^{1/3}k_f^{2/3}$, where $\epsilon$ is the energy production rate per unit mass and $k_f$ is the characteristic wave vector of the pumping force. Thus we arrive at the inequalities
 \begin{equation}
 \epsilon^{1/3}k_f^{2/3} \gg \alpha, \gamma,
 \label{weakness}
 \end{equation}
where $\alpha$ is the bottom friction coefficient, and $\gamma$ is the viscous damping rate at the pumping scale $k_f^{-1}$, $\gamma=\nu k_f^2$ ($\nu$ is the kinematic viscosity coefficient). In simulations, hyperviscosity is often used. In the case the inequalities (\ref{weakness}) are still obligatory for exciting turbulence, where $\gamma$ is the hyperviscous damping rate at the pumping scale $k_f^{-1}$.

If the inequalities (\ref{weakness}) are satisfied then turbulence is excited and pulsations of different scales are formed due to non-linear interaction of the flow fluctuations. The energy produced by the forcing at the scale $k_f^{-1}$ flows to larger scales whereas the enstrophy produced by the forcing at the same scale flows to smaller scales \cite{67Kra,68Lei,69Bat}. Thus two cascades are formed: the energy cascade (inverse cascade) realized at scales larger than the forcing scale $k_f^{-1}$ and the enstrophy cascade realized at scales smaller than the forcing scale $k_f^{-1}$. In an unbound system the energy cascade is terminated by the bottom friction at the scale
 \begin{equation}
 L_\alpha = \epsilon^{1/2}\alpha^{-3/2},
 \label{Lalpha}
 \end{equation}
where a balance between the energy flux $\epsilon$ and the bottom friction is achieved. The enstrophy cascade is terminated by viscosity (or hyperviscosity) \cite{BE12}.

If the box size $L$ is larger than $L_\alpha$ then the above two-cascade picture is realized.  We consider the opposite case $L<L_\alpha$. Then the energy, transferred to the box size $L$ by the inverse cascade, is accumulated there giving rise to a mean (coherent) flow. As numerics and experiment show, the coherent flow contains some vortices separated by a hyperbolic flow. The characteristic velocity of the coherent motion can be estimated as $\sim \sqrt{\epsilon/\alpha}$. The estimate is a simple consequence of the energy balance: the incoming energy rate $\epsilon$ has to be compensated by the bottom friction rate. The characteristic mean vorticity in the hyperbolic region is estimated as $\Omega\sim L^{-1}\sqrt{\epsilon/\alpha}$. However, inside the coherent vortices the mean vorticity $\Omega$ is much higher.

Due to the energy accumulation at the box size the average flow appears to be much stronger than the flow fluctuations. Therefore in the main approximation one can neglect fluctuations, dissipation (bottom friction and viscosity) and pumping to obtain
 \begin{equation}
 \bm V \nabla \Omega=0,
 \label{zero}
 \end{equation}
where $\bm V$ is the mean velocity and $\Omega= \mathrm{curl}\ \bm V$. The equation is valid (in the main approximation) both inside the vortices and in the hyperbolic region.

\section{Coherent Vortex}
\label{sec:vortex}

Here we examine the flow inside the coherent vortices. The average flow inside the vortex is highly isotropic (in the reference frame attached to the vortex center) \cite{Colm,Shats,14LBFKL}. Such flow can be characterized by the polar velocity $U$ dependent on the distance $r$ from the vortex center. Then the average vorticity is $\Omega=\partial_r U +U/r$. Obviously, such isotropic flow satisfies the equation (\ref{zero}). Therefore, to obtain an equation for $U$, one has to use the complete Navier-Stokes equation. Assuming that the average pumping force is zero one obtains after averaging
 \begin{equation}
 \alpha U=-\left(\partial_r+\frac{2}{r}\right)\langle uv \rangle
 +\nu \left(\partial_r^2 + \frac{1}{r}\partial_r -\frac{1}{r^2}\right) U,
 \label{Ueq}
 \end{equation}
where $v$ and $u$ are radial and polar components of the velocity fluctuations and angular brackets mean time averaging. Therefore, to find the $r$-dependence of $U$ one has to establish statistics of the flow fluctuations.

To analyze the fluctuations, it is convenient to use the equation for the fluctuating vorticity $\varpi$
 \begin{equation}
 \partial_t\varpi+(U/r)\partial_\varphi\varpi+v\partial_r\Omega
 +\nabla\left(\bm v \varpi-\langle \bm v \varpi \rangle\right)
 =\phi-\hat\Gamma \varpi.
 \label{vact}
 \end{equation}
Here $\varphi$ is polar angle, $\phi$ is curl of the pumping force, $\bm v$ is fluctuating velocity, and the operator $\hat\Gamma$ presents dissipation including both the bottom friction and the viscosity term, $\hat\Gamma=\alpha - \nu \nabla^2$. For the case of hyperviscosity the last contribution to $\hat\Gamma$ should be modified, it is substituted by $(-1)^{p+1} \nu_p (\nabla^2)^p$.

Note that in our reference system, attached to the vortex center, the fluctuating velocity $\bm v$ contains a homogeneous contribution caused by the velocity of the vortex center. The contribution produces no additional vorticity. That is why we can utilize the equation (\ref{vact}) for analyzing the flow fluctuations.

The left-hand side of the equation (\ref{vact}) changes its sign under the combined transformation
 \begin{equation}
 t\to -t, \ \varpi\to\varpi, \ \varphi\to-\varphi, \ r\to r, \
 v \to -v, \ u\to u.
 \label{transf}
 \end{equation}
Statistical properties of the pumping $\phi$ are assumed to be isotropic, invariant under $\phi\to-\phi$, and are invariant under $t\to-t$. That is why in the case $\hat\Gamma=0$ correlation functions of the velocity fluctuations have to be invariant under the transformation (\ref{transf}). Therefore the average $\langle u v \rangle$ or the average $\langle \varpi v \rangle$ should be zero, since the averages change its sign at the transformation (\ref{transf}). However, in the case $\hat\Gamma=0$ the system is inhomogeneous in time. To ensure the homogeneity, one should take a finite $\hat\Gamma$, that makes the averages non-zero. They remain finite in the limit $\hat\Gamma\to0$, it is a manifestation of the dissipation anomaly well known in turbulence. One of the manifestation of the anomaly is the so-called d'Alembert's paradox.

\subsection{Universal interval}

There is the viscous core in the center of the coherent vortex. Being interested in the region outside the core, we neglect the viscous term in the equation (\ref{Ueq}), staying with
 \begin{equation}
 \alpha U=-\left(\partial_r+\frac{2}{r}\right)\langle uv \rangle =-\langle v\varpi \rangle.
 \label{Ueq1}
 \end{equation}
The last relation in Eq. (\ref{Ueq1}) can be checked using the isotropy ($\varphi$-independence) of the averages. As it follows from Eq. (\ref{Ueq1}), the main goal of our calculations is the average $\langle uv \rangle$ or the average $\langle v\varpi \rangle$.

Further we consider the region outside the vortex core where the coherent velocity gradient is large enough,
 \begin{equation}
 U/r \gg \epsilon^{1/3}k_f^{2/3}.
 \label{passiver}
 \end{equation}
In this case fluctuations in the interval of scales between the pumping scale $k_f^{-1}$ and the radius $r$ are strongly suppressed by the coherent flow. The inequality (\ref{passiver}) means that the average velocity gradient $U/r$ is larger than the gradient of the velocity fluctuations in the interval of scales. Therefore the passive regime is realized there, that is the self-interaction of the velocity fluctuations is negligible.

Moreover, then the passive regime is realized for scales smaller than the pumping scale $k_f^{-1}$. Indeed, in the direct cascade the velocity gradients can be estimated as $\epsilon^{1/3}k_f^{2/3}$, upto logarithmic factors weakly dependent on scale, see \cite{71Kra,94FL,11FL}. Therefore the inequality (\ref{passiver}) means dominating coherent velocity gradient in the interval of scales where the direct cascade would realize.

The passive regime can be consistently analyzed. Direct calculations (see Appendix) show that
 \begin{equation}
 \langle uv \rangle  =\epsilon/\Sigma,
 \label{uvuni}
 \end{equation}
where $\Sigma$ is the local shear rate of the coherent flow
 \begin{equation}
 \Sigma=r\partial_r\left(U/r\right)
 =\partial_r U -U/r.
 \label{shear}
 \end{equation}
The expression (\ref{uvuni}) is derived at the condition $\Sigma\gg \gamma,\alpha$, that is guaranteed by the inequalities (\ref{weakness},\ref{passiver}). Substituting the expression (\ref{uvuni}) into Eq. (\ref{Ueq1}) one finds a solution
 \begin{equation}
 U=\sqrt{3\epsilon/\alpha}, \qquad
 \Sigma=-U/r,
 \label{sqrt}
 \end{equation}
for the mean profile. Thus we arrive at the flat profile of the polar velocity found in Ref. \cite{14LBFKL}.

The expression (\ref{uvuni}) is in accordance with our expectation based on symmetry reasoning. Indeed, in the absence of dissipation the average $\langle uv \rangle$ is zero due to symmetry of the system under the transformation (\ref{transf}). As we demonstrate in Appendix, the bottom friction cannot produce non-zero $\langle uv \rangle$. Therefore its value is related to viscosity (hyperviscosity) and should, therefore, be determined by short scales where the viscosity (hyperviscosity) becomes relevant and kills the flow fluctuations. (It is explicitly demonstrated in Appendix.) That is why $\Sigma$ is present in the denominator of the expression (\ref{uvuni}) since just $\Sigma$ determines the dissipation rate at the viscous scale.

The left-hand side of the inequality (\ref{passiver}) diminishes as $r$ grows. Therefore it is broken at some $r\sim R_u$. Substituting the expression (\ref{sqrt}) into Eq. (\ref{passiver}) one obtains
 \begin{equation}
 R_u = L_\alpha^{1/3}k_f^{-2/3}
 =\epsilon^{1/6}\alpha^{-1/2}k_f^{-2/3}.
 \label{scen2}
 \end{equation}
Note that $R_u$ can be larger or smaller than the box size $L$, depending on the system parameters. The case $R_u>L$ is, probably, characteristic of the numerics \cite{Colm} and the experiments \cite{Shats}, then the passive regime is realized everywhere in the box. In contrast, in numerics \cite{14LBFKL} the universal region is relatively small, $R_u<L$, and is well separated from the outer region, that is not completely passive.

\subsection{Viscous core}

The universal profile (\ref{sqrt}) implies neglecting viscosity in the equation for the average velocity. Since the average velocity (\ref{sqrt}) is independent of the separation, the mean velocity gradient increases as $r$ diminishes. In the situation the viscosity comes into game being responsible for forming the vortex core. To find the core radius $R_c$ one can use the equation for the average polar velocity (\ref{Ueq}). Comparing the left-hand side in Eq. (\ref{Ueq}) and the viscous term, one finds an estimation for the core radius
 \begin{equation}
 R_c\sim (\nu/\alpha)^{1/2}.
 \label{scen1}
 \end{equation}
At $r\ll R_c$ the viscosity dominates and therefore $U\propto r$, that corresponds to a solid rotation.

The estimate (\ref{scen1}) can be obviously generalized for the case of hyperviscosity. Taking the dissipation operator in the form $\hat\Gamma= (-1)^{p+1} \nu_p (\nabla^2)^p$, one obtains $R_c \sim (\nu_p/\alpha)^{1/(2p)}$, instead of Eq. (\ref{scen1}). If $p=1$, we return to Eq. (\ref{scen1}).

The universal profile (\ref{sqrt}) is realized in the interval of distances $R_c \ll r \ll R_u$. The interval does exist if $R_c \ll R_u$. The inequality is equivalent to inequality (\ref{weakness}) for $\gamma=\nu k_f^2$, that is $\epsilon^{1/3}k_f^{2/3}\gg \nu k_f^2$. The inequality means that the characteristic velocity gradient at the pumping scale is much larger than the viscous damping there. By other words, the inequality $R_c \ll r \ll R_u$ is equivalent to one enabling the direct cascade in the traditional two-cascade picture. The above arguments are directly generalized for the case of superviscosity.

Note that a derivation of the expression (\ref{uvuni}) implies the inequality $k_f r \gg 1$ justifying the shear approximation for the average velocity. Substituting here $r=R_c$ we find $\nu k_f^2 \gg \alpha$. More generally, the inequality $\gamma \gg \alpha$ has to be valid, that is implied in our analysis. (The inequality $\gamma \gg \alpha$ is satisfied in numerics \cite{14LBFKL}.) In the opposite case $\alpha\gg \gamma$ the shear approximation is destroyed at $r\sim k_f^{-1}$, that has to be the lower border of the profile (\ref{sqrt}). At $r$ smaller than $k_f^{-1}$ the mean velocity diminishes as $r$ decreases.

The structure of the coherent vortices in the case of zero bottom friction, $\alpha=0$, was established in \cite{15KL}. In this case just the viscosity determines the structure. 

 \subsection{Outer Region}

Let us consider the region outside the interval with the universal flat profile (\ref{sqrt}), that is the case $r>R_u$. We refer to the region as the outer one, it does exist if $R_u< L$. One expects that the average motion remains isotropic there. In the outer region $\Sigma\ll \epsilon^{1/3} k_f^{2/3}$. Therefore the passive regime is substituted here by a mixed one. In the interval of scales from $k_f^{-1}$ to $\epsilon^{1/2}/ \Sigma^{3/2}$ the traditional inverse cascade is realized whereas at larger scales the coherent motion modifies the inverse cascade essentially. The direct (enstrophy) cascade is weakly influenced by the coherent flow in the outer region.

The symmetry (\ref{transf}) leads us to the conclusion that the average $\langle uv \rangle$ is formed at the scales where the viscosity (hyperviscosity) comes into game. The property is in detail demonstrated for the passive regime where consistent calculations can be done (see Appendix). In the outer region such consistent calculations cannot be done. Therefore one should base on symmetry reasoning. Thus, we expect that the average $\langle uv \rangle$ is determined by the expression like (\ref{uvuni}) where the denominator is no other than the characteristic dissipation rate at the viscous (hyperviscous) scale. Based on this, one would expect the expression $\langle uv \rangle\sim \epsilon/(\epsilon^{1/3} k_f^{2/3})$ since $\epsilon^{1/3} k_f^{2/3}$ is just the characteristic dissipation rate in the direct cascade. (Again, to avoid a misunderstanding, note that in above reasoning we ignored a weak logarithmic dependence of the vorticity correlation functions in the direct cascade, see \cite{71Kra,94FL,11FL}.)

However, for the traditional direct cascade $\langle uv \rangle=0$ because of isotropy of the cascade. Therefore the main contribution to $\langle uv \rangle$, estimated as $\epsilon/(\epsilon^{1/3} k_f^{2/3})$, is absent. The isotropy is weakly broken by the presence of the coherent flow, its influence can be characterized by the dimensionless parameter $\Sigma/(\epsilon^{1/3} k_f^{2/3})$. One expects that the main contribution to the average $\langle uv \rangle$ is linear in $\Sigma$. Adding the factor $\Sigma/(\epsilon^{1/3} k_f^{2/3})$ to the above estimate, we find
 \begin{equation}
 \langle uv \rangle\sim
 \frac{\epsilon^{1/3}\Sigma}{k_f^{4/3}}.
 \label{uvmix}
 \end{equation}
Of course, at $r\sim R_u$ the expression (\ref{uvmix}) turns to the expression (\ref{uvuni}).

Substituting the expression (\ref{uvmix}) into Eq. (\ref{Ueq1}), one obtains
 \begin{equation}
 \alpha U \sim -\frac{\epsilon^{1/3}}{k_f^{4/3}}
 \left(\partial_r+\frac{2}{r}\right)
 \left(r\partial_r\frac{U}{r}\right).
 \label{outer1}
 \end{equation}
The expression implies a fast (exponential) decay of $U$ at $r>R_u$ on a length $\sim R_u$. Thus, $R_u$ can be treated as the coherent vortex size, where the mean vorticity is much larger than its typical value $L^{-1}\sqrt{\epsilon/\alpha}$.

One can think about the case $R_u\ll L$. Then, as we demonstrate further, a lot of vortices should appear, separated by the distance $\sim R_u$. Therefore even in the case $R_u\ll L$ there is no region outside $R_u$ where the mean flow profile is, rigorously, isotropic. Thus the consideration of this subsection is mainly qualitative. Nevertheless, the conclusion about fast attenuation of $\Omega$ at $r>R_u$ at the scale $\sim R_u$ remains correct. Note that at $R_u\ll L$ the mean vorticity at the border of the universal region, $\epsilon^{1/2} \alpha^{-1/2} R_u^{-1}$, is still much larger than the typical mean vorticity in the hyperbolic region, $\epsilon^{1/2} \alpha^{-1/2}L^{-1}$. Therefore a layer of the order $R_u$ where the vorticity diminishes from $\epsilon^{1/2} \alpha^{-1/2} R_u^{-1}$ to $\epsilon^{1/2} \alpha^{-1/2}L^{-1}$, does exist.

\section{Hyperbolic region}
\label{sec:hyperbolic}

One can generalize the symmetry reasoning formulated in Section \ref{sec:vortex} to the case of the hyperbolic region. For the purpose we introduce the curvilinear reference system related to the Lagrangian trajectories of the mean velocity $\bm V$. In the reference system the mean velocity has the only component $U$ (along the Lagrangian trajectories). Let us also introduce components of the fluctuating velocity, $u$ and $v$, longitudinal and transverse to the Lagrangian trajectories. Clearly, inside the coherent vortices the quantities pass to the ones introduced previously.

In terms of the introduced variables, the Euler equation is invariant under the transformation
 \begin{equation}
 t\to -t, \ \varpi\to\varpi, \
 \Omega \to \Omega, \
 U,u \to U,u \
 v \to -v.
 \label{flp5}
 \end{equation}
In addition, one should change sign of the coordinate along the Lagrangian trajectories. Clearly, the transformation (\ref{flp5}) is a direct generalization of the transformation (\ref{transf}) (relevant for the interior of the coherent vortices).

An invariance of the Euler equation and of the pumping statistics under the transformation (\ref{flp5}) means that the average $\langle v u \rangle$ or $\langle v \varpi \rangle$ is formed at the dissipation scale. Therefore we can use the same estimate (\ref{uvmix}) for the average $\langle u v \rangle$ for the hyperbolic region as well. Substituting there $\Sigma\sim L^{-1}\sqrt{\epsilon/\alpha}$, one obtains
 \begin{equation}
 \langle v \varpi \rangle
 \sim \sqrt{\epsilon\alpha}\,
  \frac{R_u^2}{L^2},
 \label{flp6}
 \end{equation}
where we exploited the fact that the characteristic scale of the coherent flow is $L$.

Now we examine an equation for the coherent flow that can be obtained by averaging the Navier-Stokes equation. In the principal approximation the mean velocity $\bm V$ should satisfy the equation (\ref{zero}).  We designate its solution as $\bm V_0$. Due to the dissipation terms and the nonlinear term related to the flow fluctuations there is a correction $\bm V_1$ to the velocity $\bm V_0$ that satisfies the following equation
 \begin{equation}
 \alpha\Omega_0 +(\bm V_0 \nabla)\Omega_1
 +(\bm V_1 \nabla)\Omega_0 +\nabla \langle \bm v \varpi \rangle=0.
 \label{flp2}
 \end{equation}
Here we omitted the dissipation term (that is irrelevant outside the vortex cores) and taken into account that the average force (exciting the turbulence) is equal to zero.

Multiplying the equation (\ref{flp2}) by $\Omega_0$ and integrating the result over the whole box one finds
 \begin{equation}
 \alpha \int dS\ \Omega_0^2
 -\int dS\ \nabla\Omega_0 \langle \bm v \varpi \rangle=0.
 \label{flp3}
 \end{equation}
Here we utilized the equation (\ref{zero}), $\bm V_0 \nabla \Omega_0=0$, the incompressibility condition and the boundary conditions. The relation (\ref{flp3}) is correct both for the periodic setup and for the case of a box with zero velocity at the boundaries. It is a manifestation of the existence of a zero mode of the equation  (\ref{zero}), that can be obtained by multiplying $\Omega_0$ (or $\bm V_0$) by a constant. As it follows from Eq.  (\ref{zero}), $\nabla\Omega_0$ is perpendicular to $\bm V_0$. Therefore one can rewrite the equation (\ref{flp3}) as
 \begin{equation}
 \alpha \int dS\ \Omega_0^2=
 \int dS\ |\nabla\Omega_0| \langle v \varpi \rangle.
 \label{flp4}
 \end{equation}
It is satisfied in the main approximation in the fluctuations weakness.

Now we can substitute into the relation (\ref{flp4}) the estimates $\Omega_0\sim L^{-1}\sqrt{\epsilon/\alpha}$ and $|\nabla\Omega_0|\sim L^{-2}\sqrt{\epsilon/\alpha}$. Then we conclude that the relation cannot be satisfied if $R_u \ll L$. Therefore we expect that in the case $R_u\ll L$ a number of the coherent vortices should appear in the box separated by a distance $\sim R_u$. That is the only possibility, that we see, to overcome the discrepancy associated with the relation (\ref{flp4}) and the estimate (\ref{flp6}).

\section{Conclusion}

We investigated analytically the coherent flow generated by the inverse cascade in a restricted box that consists of a number of vortices and a hyperbolic flow between them. The mean velocity can be estimated as $\sqrt{\epsilon/\alpha}$ (where $\epsilon$ is the energy production rate and $\alpha$ is the bottom friction coefficient) everywhere. Besides, the mean vorticity inside the vortices is much larger than in the hyperbolic region, that can be estimated as $\sqrt{\epsilon/\alpha}/L$. The flow inside the vortices is complicated. It can be divided into some regions: the viscous core, the universal interval and the outer region. In the universal interval the velocity fluctuations are passive and the polar mean velocity of the flow is characterized by the flat profile $U=\sqrt{3 \epsilon/\alpha}$.

We established the values of the viscous core radius $R_c$ (\ref{scen1}) and of the vortex size $R_u$ (\ref{scen2}). Subsequent analysis shows that the coherent vortex vorticity diminishes fast in the outer region $r>R_u$, the characteristic length of this decay is $\sim R_u$. Our analysis of the outer vortex region $r>R_u$ is based on symmetry reasoning. The reasoning are confirmed by consistent calculations performed for the passive regime realized at $r<R_u$. A relation between the box size $L$ and the vortex radius $R_u$ can be arbitrary. The case $R_u>L$ is realized in the numerics \cite{Colm} and experiments \cite{Shats}, then the vortex is not clearly separated from the hyperbolic region. In this case the passive regime is realized everywhere. In contrast, in numerics \cite{14LBFKL} the universal region is relatively small, $R_u<L$, and is well separated from the hyperbolic region, where the fluctuations are not completely passive.

In numerics \cite{Colm,14LBFKL}, dealing with the periodic setup, two coherent vortices were observed, in experiments \cite{Shats} a few vortices were observed as well. In this case the hyperbolic coherent motion is characterized by a scale determined by the box size $L$. Then vortices are placed in the stagnation points of the hyperbolic flow (upto fluctuations). However, in the the limit $R_u\ll L$ we expect appearing a lot of vortices with complicated hyperbolic flow between them, see Section \ref{sec:hyperbolic}. Most probably they will arrange as a lattice. A natural space structure for the lattice is a chessboard. (Similar lattice was observed in \cite{Smith2}.) However, other possibilities (say, the hexagonal or the honeycomb structure) are not excluded.

We formulated conditions at which the coherent vortices with the universal profile (\ref{sqrt}) appear in $2d$ turbulence in a finite box. One of the conditions is that the viscous (hyperviscous) dissipation at the pumping scale $k_f^{-1}$ should be much weaker than the characteristic velocity gradient $\epsilon^{1/3} k_f^{2/3}$. This condition was violated in early simulations \cite{Smith1,Smith2,Borue} (to extend the interval for the inverse cascade). If $\epsilon^{1/3} k_f^{2/3}$ is of the order of the viscous damping at the pumping scale then $R_u\sim R_c$ and the universal interval is absent. That is why the universal profile (\ref{sqrt}) was not observed in the works.

Note also that the inverse energy cascade is observed for surface solenoidal turbulence excited by waves caused by Faraday instability \cite{Kameke,Xia2}. It would be interesting to extend our analysis to this case. It is a subject of future investigations.

\acknowledgements

We thank valuable discussions with G. Boffetta and G. Falkovich. The work is supported by RScF grant 14-22-00259.

 \appendix

 \section{Passive regime}

Here we investigate the passive regime of flow fluctuations on the background of the coherent isotropic vortex characterized by the average polar velocity $U(r)$. In this case consistent calculations can be done. As we already noted in main body, the passiveness implies the inequality (\ref{passiver}) that guarantees weakness of the interaction of the flow fluctuations at all scales. Neglecting the non-linear terms (responsible for the interaction) in the equation (\ref{vact}) one obtains the linear equation for the fluctuating vorticity
 \begin{equation}
 \partial_t\varpi+(U/r)\partial_\varphi\varpi+v\partial_r\Omega
 =\phi-\hat{\Gamma}\varpi.
 \label{vpas}
 \end{equation}
The equation (\ref{vpas}) describes dynamics of the flow fluctuations on the background of the average (coherent) flow.

The dissipation in Eq. (\ref{vpas}) (the last term in the equation) is caused both by the bottom friction and the viscosity (or, in a numerics, by the hyperviscosity). Therefore in Fourier representation the operator $\hat\Gamma$ can be written as
 \begin{equation}
 \Gamma(k)=\alpha  + \gamma (k/k_f)^{2p},
 \label{dissip}
 \end{equation}
where $k$ is a wave vector and $k_f$ is the characteristic wave vector of the pumping force. For the viscosity $p=1$, for the hyperviscosity $p>1$. Due to the inequalities (\ref{weakness},\ref{passiver}) the inequality $\Sigma \gg \alpha,\gamma$ is satisfied, where $\Sigma$ is the local shear rate of the coherent flow defined by Eq. (\ref{shear}).

Below we assume the inequality $k_f r \gg 1$. As we explained in the main body, in whole region of the existence of the universal profile (\ref{sqrt}), $R_c<r<R_u$, the condition $k_f r \gg 1$ is guaranteed by the inequality $\alpha \ll \gamma$. The smallness of the pumping scale $k_f^{-1}$ in comparison with the radius $r$ enables one to pass to the shear approximation for the vorticity fluctuations that is correct in the leading order in $(k_f r)^{-1}$. The effective shear rate is determined by Eq. (\ref{shear}).

Let us consider the fluctuation dynamics near a circle of the radius $r_0$. Passing to the reference system rotating with the angular velocity $\Omega_0$ (corresponding to the radius $r_0$), one finds from Eq. (\ref{vpas}) in the leading order
 \begin{equation}
 \partial_t\varpi+\Sigma x_1\frac{\partial\varpi}{\partial x_2}
 +\hat{\Gamma}\varpi=\phi,
 \label{vispas}
 \end{equation}
where $\Sigma$ is taken at $r=r_0$, $x_1=r-r_0$ is the radial coordinate, and $x_2=r_0\varphi$ is the angular coordinate. The equation (\ref{vispas}) describes the passive evolution of the flow fluctuations in the shear flow with the shear rate $\Sigma$.

The equation (\ref{vispas}) leads to homogeneous statistical properties of the flow fluctuations in the space $(x_1,x_2)$. That is why it is worth to make Fourier-transform over $x_1$ and $x_2$. Rewriting the evolution equation (\ref{vispas}) for the spatial Fourier component of the vorticity $\varpi_{\bm{k}}$, one obtains
 \begin{equation}
 \partial_t\varpi_{\bm{k}}-\Sigma k_2\partial\varpi_{\bm{k}}/\partial k_1
 +\Gamma(k)\varpi_{\bm{k}}=\phi_{\bm k}.
 \label{pass1}
 \end{equation}
A formal solution of the equation (\ref{pass1}) is written as
 \begin{eqnarray}
 \varpi_{\bm{k}}(t)
 =\int^t d\tau\, \phi[\tau,k_1+(t-\tau)\Sigma k_2,k_2] \qquad
 \label{passolv} \\
 \times\exp\left\{-\int\limits_\tau^t\,d\tau'\Gamma
 \left(\sqrt{\left[k_1+(t-\tau')\Sigma k_2\right]^2+k_2^2}\right)\right\},
 \nonumber
 \end{eqnarray}
where the integral is taken over the time interval where the pumping is switched on.

Further we assume that the pumping is short correlated in time. Then it is characterized by the pair correlation function that can be written as
 \begin{equation}
 \langle \phi_{\bm{k}}(t) \phi_{\bm{q}}(t')\rangle
 =2(2\pi)^2\epsilon\delta(\bm{k}+\bm{q})\delta(t-t')k^2\chi(\bm{k}),
 \label{pump}
 \end{equation}
in the Fourier representation. The function $\chi(\bm{k})$ is concentrated in a vicinity of $k_f$, it has to be normalized as
 \begin{equation}
 \int\frac{d^2\bm{k}}{(2\pi)^2}\chi(\bm{k})=1,
 \label{normachi}
 \end{equation}
to provide the energy pumping rate $\epsilon$. We also assume isotropy of pumping, that is the function $\chi(\bm k)$ depends solely on $k$, that is the absolute value of $\bm k$.

One directly finds from the expressions (\ref{passolv},\ref{pump}) the simultaneous pair correlation function of the vorticity
 \begin{eqnarray}
 \langle\varpi_{\bm{k}}(t)\varpi_{\bm{q}}(t)\rangle
 =2(2\pi)^2\epsilon \delta(\bm{k}+\bm{q})
 \nonumber \\ \times
 \int\limits_0^T d\tau\, \left[\left(k_1+\Sigma\tau k_2\right)^2+k_2^2\right]^2
 \chi\left(k_1+\Sigma\tau k_2,k_2\right)
 \label{corr1} \\
 \times
 \exp\left\{-2\int\limits_0^\tau d\tau'\,\Gamma
 \left(\sqrt{\left[k_1+\Sigma\tau'k_2\right]^2+k_2^2}\right)\right\}.
 \nonumber
 \end{eqnarray}
where $T$ is a duration of the period before $t$ when the pumping was switched on. In the stationary case one should take the limit $T\to\infty$. We are interested just in this stationary case.

Using the relation $v_{\bm{k}}=i\epsilon_{\alpha\beta}(k_\beta/k^2)\varpi_{\bm{k}}$ we can calculate correlation functions of the velocity fluctuations in the Fourier representation. Making then the inverse Fourier transform, we find the velocity correlation function. Below, we concentrate on the single-point correlation function $\langle u v \rangle$, that is written as
 \begin{eqnarray}
 \langle uv \rangle = - 2\epsilon \int_0^T d\tau
 \int \frac{d^2 k}{(2\pi)^2}
 \frac{k_1 k_2}{(k_1^2+k_2^2)^2}
 \nonumber \\
 \left[\left(k_1+\Sigma\tau k_2\right)^2+k_2^2\right]^2
 \chi\left(k_1+\Sigma\tau k_2,k_2\right)
 \nonumber \\
 \times
 \exp\left\{-2\int\limits_0^\tau d\tau'\,\Gamma
 \left(\sqrt{\left[k_1+\Sigma\tau'k_2\right]^2+k_2^2}\right)\right\}.
 \label{uvint}
 \end{eqnarray}
Further we pass to the variable $\bm q=(k_1+\Sigma\tau k_2,k_2)$.

We first consider the case $\Gamma(k)=0$. Then the passive equation (\ref{vpas}) is invariant under the transformation (\ref{transf}) as well as the complete equation (\ref{vact}). That is why the integral
 \begin{equation}
 -2\epsilon\int\limits_0^{T} d\tau\, \int \frac{d^2{q}\,q^2
 \chi(\bm{q})}{(2\pi)^2}
 \frac{(q_1-\Sigma\tau q_2)q_2}{\left[(q_1-\Sigma\tau q_2)^2+q_2^2\right]^2},
 \label{passive0}
 \end{equation}
determining $\langle u v \rangle$ at $\Gamma=0$, is zero at any finite $T$. Indeed, the average $\langle u v \rangle$ changes its sign at the transformation (\ref{transf}) and should be equal to zero at $\Gamma=0$. Introducing a finite $\Gamma$ brakes the symmetry and makes the average $\langle uv \rangle$ non-zero.

Further we analyze the stationary case (implying $\Gamma\neq 0$), then we take the limit $T\to\infty$. Passing to the variable $\bm q=(k_1+\Sigma\tau k_2,k_2)$ and taking an integral in part one finds from Eq. (\ref{uvint})
 \begin{equation}
 \langle uv \rangle \equiv \langle v_1 v_2\rangle
 =\frac{\epsilon}{\Sigma}(1-Q),
 \label{uv3}
 \end{equation}
where $Q$ is defined as
 \begin{eqnarray}
 Q=2\int\limits_0^\infty d\tau\, \int \frac{d^2\bm{q}\,q^2\chi(\bm{q})}{(2\pi)^2}
 \frac{\Gamma\left(\sqrt{\left(q_1-\Sigma\tau q_2\right)^2+q_2^2}\right)}
 {(q_1-\Sigma\tau q_2)^2+q_2^2}
 \nonumber \\
 \times \exp\left[-2\int\limits_0^\tau d\tau'\,\Gamma
 \left(\sqrt{\left(q_1-\Sigma\tau'q_2\right)^2+q_2^2}\right)\right]. \qquad
 \label{uvQ}
 \end{eqnarray}
The expression is a starting point of the subsequent analysis.

It is instructive to consider the case $\Gamma=\alpha$, that is $\gamma=0$. In this case $Q=1$. It can be explicitly obtained from the expression (\ref{uvQ}) after integrating over the polar angle in the $q$-space and using the condition (\ref{normachi}). Let us stress that this property exploits isotropy of the pumping statistics, that is the assumption that $\chi$ is a function of the absolute value of $\bm q$. Thus, in this case $\langle uv \rangle =0$. Therefore, though the bottom friction breaks the symmetry under the transformation (\ref{transf}), it cannot produce a nonzero value of $\langle uv \rangle =0$.

Now we analyze the opposite case where viscosity (hyperviscosity) is stronger than the bottom friction at the pumping scale $k_f^{-1}$, $\gamma\gg \alpha$. As we explained in the main body, just this case is relevant. Then $\alpha$ can be neglected in the expression (\ref{uvQ}) and we stay with $\Gamma(k)=\gamma (k/k_f)^{2p}$.

An inspection of the expression (\ref{uvQ}) shows that for $p>1/2$ both $q_1$ and $q_2$ are of order of $k_f$. The estimate implies that $\chi(\bm q)$ decreases fast enough as $q$ grows. There is a potentially dominant contribution to the value of $Q$ from the region $q_2\ll k_f$. However, at $p>1/2$ the contribution gaining from small $q_2$ is smaller than the one gaining from $q_2\sim k_f$. If $q_1,q_2\sim k_f$ then for a characteristic $\tau$ the inequality $\Sigma\tau\gg1$ is satisfied. Then we easily obtain the estimate
 \begin{equation}
 Q\sim \left( \frac{\gamma}{\Sigma}
 \right)^{2/(2p+1)}\ll1.
 \label{estQ}
 \end{equation}
Thus, we arrive at the conclusion that if viscosity (or hyperviscosity) dominates over $\alpha$ at the pumping scale $k_f^{-1}$ then $Q\ll1$ in the passive regime. Therefore the expression (\ref{uv3}) leads to the expression (\ref{uvuni}).

Note, that the characteristic time $\tau_0$ in the expression (\ref{uvint}) or in the expression (\ref{uvQ}) is determined by the condition $\tau_0 \Gamma(k_f \Sigma \tau_0)\sim 1$, that is $\gamma\tau_0 (\Sigma \tau_0)^{2p}\sim 1$. The relation determines the characteristic time that is needed to enhance (by the influence of the coherent flow) the wave vector $\bm q$ from the initial value $\sim k_f$ to the value $\sim \tau_0 \Sigma k_f$. The enhancement is caused by the deformation of the flow fluctuations (produced by pumping) in the coherent shear flow. At $\tau\sim \tau_0$ the dissipation comes into game and kills the flow fluctuations. We conclude that the main contribution to the average $\langle uv \rangle$ is determined by small scales (large wave vectors) where dissipation is relevant. This conclusion is in accordance with our symmetry arguments based on the transformation (\ref{transf}).

\end{document}